\documentclass[default,iicol]{sn-jnl}

\usepackage{graphicx}%
\usepackage{multirow}%
\usepackage{amsmath,amssymb,amsfonts}%
\usepackage{amsthm}%
\usepackage{mathrsfs}%
\usepackage[title]{appendix}%
\usepackage{xcolor}%
\usepackage{textcomp}%
\usepackage{manyfoot}%
\usepackage{booktabs}%
\usepackage{algorithm}%
\usepackage{algorithmicx}%
\usepackage{algpseudocode}%
\usepackage{listings}%

\usepackage{physics}

\theoremstyle{thmstyleone}%
\theoremstyle{thmstyletwo}%

\theoremstyle{thmstylethree}%

\begin{document}

\title[A potential approach to the $X(3872)$ thermal behavior]{A potential approach to the $X(3872)$ thermal behavior}


\author[1]{\fnm{N\'estor} \sur{Armesto}}\email{nestor.armesto@usc.es}

\author[2]{\fnm{Miguel \'Angel} \sur{Escobedo}}\email{miguel.a.escobedo@fqa.ub.edu}

\author[1]{\fnm{Elena G.} \sur{Ferreiro}}\email{elena.gonzalez.ferreiro@usc.es}

\author*[1]{\fnm{V\'{\i}ctor} \sur{L\'opez-Pardo}}\email{victorlopez.pardo@usc.es}

\affil[1]{\orgdiv{Instituto Galego de F\'{\i}sica de Altas Enerx\'{\i}as IGFAE}, \orgname{Universidade de Santiago de Compostela}, \city{Santiago de Compostela}, \postcode{15782}, \state{Galicia}, \country{Spain}}

\affil[2]{\orgdiv{Departament de F\'{\i}sica Qu\`antica i Astrof\'{\i}sica and Institut de Ci\`encies del Cosmos}, \orgname{Universitat de Barcelona}, \orgaddress{\street{Mart\'{\i} i Franqu\`es 1}, \city{Barcelona}, \postcode{08028}, \state{Catalonia}, \country{Spain}}}


\abstract{
We study the potential of $X(3872)$ at finite temperature in the Born-Oppenheimer approximation under the assumption that it is a tetraquark. We argue that, at large number of colors, it is a good approximation to assume that the potential consists in a real part plus a constant imaginary term. The real part is then computed adapting an approach by Rothkopf and Lafferty and using as input lattice QCD determinations of the potential for hybrids. This model allows us to qualitatively estimate at which temperature range the formation of a heavy tetraquark is possible, and to propose a qualitative picture for the dissociation of the state in a medium. Our approach can be applied to other suggested internal structures for the $X(3872)$ and to other exotic states.
}




\maketitle

\section{Introduction}\label{sec0}
Quarkonium states have a prominent role in the study of Quantum Chromodynamics (QCD), due to the high scale given by the large mass of the heavy quarks. The properties of production and absorption of quarkonium in nuclear collisions provide quantitative inputs for the study of QCD at high density and temperature~\cite{Andronic:2015wma}. 
In fact, due to its heavy mass, quarkonium production off nuclei is one of the most exploited probes for studying properties of the matter created in ultrarelativistic heavy-ion collisions, such as the strength of interactions and possible thermalization. Moreover, according to lattice QCD calculations, at sufficiently large energy densities hadronic matter undergoes a phase transition to a plasma of deconfined quarks and gluons (quark-gluon plasma, QGP). Thus, the study of quarkonium production and suppression is among the most compelling investigations in this field since the QCD binding potential is expected to be screened in the QGP phase.

 In the recent years, several quarkonium species that are difficult to accommodate in the standard quark model and in the potential model approach have been discovered, see, e.g.,~\cite{Brambilla:2022ura} and references therein. 
 The experimental observation of charmonium-like and bottomonium-like states that cannot be identified with the traditional mesons and baryons has motivated different theoretical interpretations such as molecule and tetraquark models.
 Among these exotic hadrons, the $X(3872)$, also known as $\chi_{c1}(3872)$, plays a prominent role since it was the first to be discovered~\cite{Belle:2003nnu} more than two decades ago.
 Recently, it has been detected in heavy-ion collisions at the Large Hadron Collider (LHC)~\cite{CMS:2021znk}, offering the opportunity to study exotic particles in the medium. In fact, the behavior of exotic states in a QGP is an additional piece of information that may not only help to characterize the medium properties but also provide insight about their internal structure~\cite{Wu:2020zbx,Chen:2021akx,Yun:2022evm,MartinezTorres:2014son}.

While there is an intense debate about the nature, tetraquark or molecule, of the $X(3872)$ (see, e.g., \cite{Esposito:2016noz}), in this work we adhere to the tetraquark interpretation because it allows the use of approximations, like the Born-Oppenheimer one, previously employed in the study of heavy-light systems, and exploit the analogy of the tetraquark state to quarkonia. We reduce the four-body problem to a effective two-body one, where the light degrees of freedom are integrated out and we can consider the potential between the $c$ and the $\bar c$. This simplification allows the use of the Born-Oppenheimer approximation, of lattice results that have been computed under this approximation, and of techniques and extensions of potential models previously developed for quarkonium states. Our framework, in which we consider that the separation between the $c$ and the $\bar c$ to be small and we consider the $c\bar c$ pair to be in a single color state (octet), could be extended to the molecular interpretation, a line of research that will be explored in the future.

It is our purpose here to study the thermal properties of the $X(3872)$ in medium and
to qualitatively estimate the temperature range for its formation. 
To do so, we will develop the potential of $X(3872)$ at finite temperature in the Born-Oppenheimer approximation under the assumption that it is a tetraquark. We will include the real and imaginary parts of the potential. For the real part, we will use a lattice-inspired potential for hybrids and we will follow the approach of Rothkopf and Lafferty~\cite{Lafferty:2019jpr}, while we will show that the imaginary part behaves as constant. We will discuss the survival probability of the $X(3872)$ and its time dependence. We will finish with our conclusions and future lines of work. 

The manuscript is organized as follows: in Section~\ref{sec1} we develop our potential model for vacuum and its extension to finite temperatures, and discuss the specific form that we adopt for the $X(3872)$. In Section~\ref{sec2} we present our results on the solution of the Schr\"odinger equation and the survival probability. Finally, in Section~\ref{sec3} we provide our conclusions.

\section{Potential model}\label{sec1}

\subsection{Vacuum case}\label{sec1-1}

Quarkonium mesons can be accurately described using potential models~\cite{Brambilla:2010cs}.
This approach can also be applied to the internal structure of exotic quarkonium states, as for example quarkonium hybrids and tetraquarks. In the same way that conventional quarkonium can be seen as a QCD analogue of the hydrogen atom, heavy hybrids and tetraquarks would be analogues to some electromagnetic molecules. Exotic quarkonium can be treated in the Born-Oppenheimer approximation when two conditions are met~\cite{Soto:2020xpm}. The first one is that the non-relativistic approximation is fulfilled, i.e., the velocity of the heavy quarks around the center-of-mass of the meson, $v$, must be small. The second condition is that the binding energy (of the order of $m_Q v^2$ for a non-relativistic state) must be much smaller than the QCD scale $\Lambda_\mathrm{QCD}$. To continue with this analogy, the heavy quarks will take the role of the nuclei as slow degrees of freedom while light quarks and gluons will act as the fast degrees of freedom. Then, the net effect of the light degrees of freedom is encoded in their collective quantum numbers and a potential given by their energy in the limit of heavy quarks with infinite mass, so they act as static sources.

While in the following we provide arguments for our model for the $X(3872)$, other models exist beyond those that assume a molecular state. For example, diquark-antidiquark models~\cite{Maiani:2004vq,Padmanath:2015era}, where $cq$ and $\bar{c}\bar{q}$ act as effective degrees of freedom, have been applied to this state. We also make the simplification of considering a single color channel, the octet, which we take as a reasonable assumption for the small distances that we will consider where one-gluon exchange is expected to dominate the potential. This assumption also allows the use of lattice results for hybrids composed of heavy quarks and gluons, which are the states closest to the $X(3872)$ for which results for the potential under the Born-Oppenheimer approximation are available. We note that taking for granted all the aforementioned, the nature of the degrees of freedom for the description is irrelevant. While we could apply our assumptions for other degrees of freedom, we find more justified to consider them to be heavy quarks in order to apply the existing lattice results for hybrids.

As previously mentioned, we will study the case in which the $X(3872)$ is a tetraquark. In principle, we should select which quantum numbers for the light degrees of freedom are compatible with the observed degrees of freedom of $X(3872)$ and use a lattice QCD determination of the corresponding potentials. Unfortunately, those potentials have not been computed yet. However, for a first qualitative study of thermal effects, we will assume the same parametric behavior that has been observed for hybrids: On one hand, there is more lattice data available on hybrid potentials since they can be computed in the quenched approximation. On the other hand, we do not expect the parametric behavior to be much different.

Let us now summarize our model for the $T=0$ potential of the tetraquark:
\begin{itemize}
    \item When the separation between the heavy quarks, $r$, is small, the potential can be computed in perturbation theory except for a constant given by the corresponding gluelump mass (more precisely, the tetraquark equivalent of a gluelump)~\cite{Bali:2003jq}. Therefore, the potential at short distances is Coulombic.
    \item At intermediate distances, it has been observed that the hybrid potential rises like $r^2$~\cite{Capitani:2018rox,Schlosser:2021wnr}. We assume that this is also the case for the tetraquark state.
    \item At large distances the potential rises linearly with the same string tension as the singlet potential: When $\Lambda_\mathrm{QCD}r\gg 1$, effective string theory becomes applicable \cite{Luscher:2004ib}. The string-like flux tube observed in QCD at large distances neutralizes the color of the heavy quarks and, therefore, the potential is the same in the singlet and in the octet channel.
    \item In principle, we should consider that the potential is different depending on the quantum numbers of the light degrees of freedom. However, given that the qualitative features are expected to be similar, as observed on the lattice~\cite{Capitani:2018rox,Schlosser:2021wnr}, we are going to neglect this difference.
\end{itemize}
Taking all this into account, we will use the following parametrization of the potential at $T=0$ and at small and intermediate distances:
\begin{equation}
V(r,0)=\frac{A_{-1}}{r}+A_0+A_2r^2\ .\label{V0}
\end{equation}
We will also consider that at large distances it is equal to the singlet potential, but we will not need to use an explicit formula. In particular, this implies that at large distances the potential rises linearly, although this behavior is not captured by the previous formula. However, eq. (\ref{V0}) describes accurately the regions that we need in order to obtain the binding energy and the wave function of the first bound state.

\subsection{Extension to finite $T$}\label{sec1-2}
Now, let us discuss how we could extend this potential at finite temperature. We expect it to develop some kind of screening and an imaginary part. 
The existence of an imaginary part of the potential for quarkonium at finite temperature is a well-established fact both in perturbation theory \cite{Laine:2006ns} and on the lattice \cite{Bala:2021fkm}. Extending this reasoning  to a tetraquark in the Born-Oppenheimer approximation, we make the following observations about the imaginary part of the tetraquark potential:
\begin{itemize}
    \item At small $r$, perturbation theory is valid. The computation would be similar to the one in~\cite{Laine:2006ns}, just changing the color factors. In the case of the singlet, the imaginary potential vanishes in the $r\to 0$ limit because the medium can not resolve the bound state. However, in the case of a tetraquark, the system formed by the heavy quarks has a net color charge, therefore the imaginary part of the potential goes to a constant as $r\to 0$.
    \item At large distances, the heavy quark and  antiquark do not interact with each other. Therefore, the imaginary potential should be a constant. This is observed in the perturbative computation of~\cite{Laine:2006ns}. For the same reason, in this limit the imaginary part of the potential should not depend on the color state of the heavy quarks.
    \item Taking the imaginary part of two heavy quarks in a color octet state to be a constant can also be justified in the limit of large number of colors $N_c$. The heavy quark and antiquark behave in this limit as two uncorrelated particles~\cite{Escobedo:2020tuc}.
\end{itemize}
All these observations point to the same direction. We can consider that the imaginary part of the tetraquark potential is approximated by a constant whose size is of the order of the imaginary part of the singlet potential at large distances.

Concerning the extension of the real part of the potential to finite temperature, we use the approach developed in~\cite{Lafferty:2019jpr}, which is able to describe reasonably well the real part of the singlet potential computed on the lattice at finite temperature taking as input the potential at $T=0$. The vacuum potential gets modified due to the presence of a medium. In linear response theory, the potential is modulated through the multiplication by the inverse of the static dielectric
constant or permittivity in momentum space.
Given this, the relationship between the vacuum potential, $V_{\mathrm{vac}}(\mathbf{p})$, and its in-medium counterpart, $V(\mathbf{p})$, reads
\begin{equation}
    V(\mathbf{p})=\frac{V_{\mathrm{vac}}(\mathbf{p})}{\varepsilon(\mathbf{p},m_D)},\label{Vp}
\end{equation}
where $\varepsilon(\mathbf{p},m_D)$ represents the permittivity of the medium, with $m_D$ the Debye or \mbox{screening} mass which is directly related to the temperature of the medium and vanishes for $T=0$. The inverse of the Debye mass, called Debye length, accounts for the size of the charge carrier's range of action. The permittivity is defined as an appropriate limit of the real-time in-medium gluon propagator, and its role here is to imprint the effects of the medium onto the potential.

In coordinate space, the relation~\eqref{Vp} can be written as a Fourier convolution, which will be denoted by a star:
\begin{equation}
    V(\mathbf{r},m_D)=(V_{\mathrm{vac}}*\varepsilon^{-1})(\mathbf{r},m_D).\label{VFou}
\end{equation}

The most straightforward approach would be performing the convolution, but it can be a bit cumbersome.
So, instead of doing the Fourier convolution, we will generate a differential equation that the in-medium potential verifies. Its solution gives the desired potential and we fix the integration constants by forcing the in-medium potential to smoothly match the vacuum case as the temperature goes to zero. 

We generalize the Gauss's law for a point particle of unit charge,
\begin{equation}
    \nabla\cdot\left(\frac{\mathbf{\hat{r}}}{r^2}\right)=4\pi\delta(\mathbf{r}),
\end{equation}
where $\mathbf{\hat{r}}/r^2$ is the (chromo)electric field, to a general radial field $\mathbf{E}(r)=E(r)\mathbf{\hat{r}}$. Then, in the vacuum we trivially get the equation
\begin{equation}
    \nabla\cdot\left(\frac{\mathbf{E}}{r^2E(r)}\right)=4\pi\delta(\mathbf{r}).
\end{equation}

In terms of the vacuum potential $V_{\mathrm{vac}}(r)$ the chromoelectric field is $\mathbf{E}(r)=-\nabla V_{\mathrm{vac}}(r)$, so the previous equation can be written as
\begin{align}\label{eq6}
    \left(\frac{1}{r^2E(r)^2}\pdv{E(r)}{r}\pdv{}{r}-\frac{1}{r^2E(r)}\pdv[2]{}{r}\right)V_{\mathrm{vac}}(r)\nonumber\\=\mathcal{G}(r)V_{\mathrm{vac}}(r)=4\pi\delta(\mathbf{r}).
\end{align}
Applying the linear differential operator $\mathcal{G}$ to~\eqref{VFou} and using~\eqref{eq6}, we obtain
\begin{equation}
    \mathcal{G}(r)V(r,m_D)=4\pi\varepsilon^{-1}(r,m_D),
\end{equation}
so we get
\begin{equation}
    \frac{1}{r^2}\pdv{r}\left(\frac{-1}{E(r)}\pdv{V(r,m_D)}{r}\right)=4\pi\varepsilon^{-1}(r,m_D),
\end{equation}
which formally solves as
\begin{multline}
    V(r,m_D)=C-b\int^r\dd r'E(r')\\-4\pi\int^r\dd r'E(r')\int^{r'}\dd r''r''^2\varepsilon^{-1}(r'',m_D).
\end{multline}
Next, we substitute $E(r)=-\pdv*{V_{\mathrm{vac}}(r)}{r}$, getting
\begin{multline}
    V(r,m_D)=C+bV_{\mathrm{vac}}(r)\\+4\pi\int^r\dd r'\pdv{V_{\mathrm{vac}}(r')}{r'}\int^{r'}\dd r''r''^2\varepsilon^{-1}(r'',m_D).
\end{multline}

The last step is to fix the constants $b$ and $C$ by forcing a smooth transition between $V(r,m_D)$ and $V^{\mathrm{vac}}(r)$ as $m_D\to0$. 

To determine the value of $b$ we use the fact that $V_{\mathrm{vac}}(r)=V(r,0)$ and that $\varepsilon^{-1}(r,0)$ corresponds to the Dirac delta function, obtaining
\begin{equation}
    V(r,0)
    =C+(b+1)V_{\mathrm{vac}}(r).
\end{equation}
Imposing  $V(r,0)=V_{\mathrm{vac}}(r)$, we get $b=0$, $C=0$ for $m_D=0$. As $b$ is a dimensionless constant, it cannot depend on $m_D$ because there is no other dimensionful quantity in the problem, so $b=0$ for all $m_D$.

In order to determine $C$ we can use the fact that the derivative with respect to $m_D$
must be continuous with $V_{\mathrm{vac}}(r)$.
The general formula then reads
\begin{eqnarray}
    &&V(r,m_D)=C(m_D)\\
    &&+4\pi\int^r\dd r'\pdv{V(r',0)}{r'}\int^{r'}\dd r''r''^2\varepsilon^{-1}(r'',m_D),\nonumber
\end{eqnarray}
with the prescription that $C(m_D)$ must ensure that $V(r,m_D)=V_{\mathrm{vac}}(r)$ up to $\mathcal{O}(m_D)$.

To proceed further we need to specify the permittivity. Following~\cite{Lafferty:2019jpr}, we use the HTL resummation with a Debye mass fitted to reproduce lattice data on the real part of the potential. While it does not apply at large distances, we are motivated by its success in the singlet sector.
The HTL permittivity in momentum space reads
\begin{equation}
    \varepsilon^{-1}(p,m_D)=\frac{p^2}{p^2+m_D^2}-i\pi T\frac{pm_D^2}{(p^2+m_D^2)^2},
\label{eq:permitivity}
\end{equation}
which in position space translates into
\begin{multline}
    \varepsilon^{-1}(r,m_D)=\frac{\delta(r)}{4\pi r^2}-\frac{m_D^2e^{-m_Dr}}{4\pi r}\\-i\frac{m_DT}{4\sqrt{\pi}r}G^{2,1}_{1,3}\left(\left.\begin{array}{c}-\frac{1}{2}\\\frac{1}{2},\frac{1}{2},0\end{array}\right|\frac{1}{4}m_D^2r^2\right),
\end{multline}
where $G$ denotes Meijer G-functions. Since
\begin{multline}
    4\pi\int^{r'}\dd r''r''^2\varepsilon^{-1}(r'',m_D)=e^{-m_Dr'} (m_Dr'+1)\\-i \frac{\sqrt{\pi }m_DT}{2} r'^2G^{2,1}_{1,3}\left(\left.\begin{array}{c}\frac{1}{2}\\\frac{1}{2},\frac{1}{2},-1\end{array}\right|\frac{1}{4}m_D^2r'^2\right),
\end{multline}
we get\footnote{Although we will not use it, we show here the imaginary part for completeness.}
\begin{multline}
\Re[V(r,m_D)]=\Re[C(m_D)]\\+\int^r\dd r'\pdv{V(r',0)}{r'}e^{-m_Dr'} (m_Dr'+1)
\end{multline}
and
\begin{multline}
    \Im[V(r,m_D)]=\Im[C(m_D)]\\
-\int^r\dd r'\pdv{V(r',0)}{r'}\frac{\sqrt{\pi }m_DT}{2} r'^2\\\times G^{2,1}_{1,3}\left(\left.\begin{array}{c}\frac{1}{2}\\\frac{1}{2},\frac{1}{2},-1\end{array}\right|\frac{1}{4}m_D^2r'^2\right).\label{imVV}
\end{multline}
Assuming that $C(m_D)$ is real 
since the integral in~\eqref{imVV} is $\mathcal{O}(m_D^2)$, we get
\begin{multline}
\Re[V(r,m_D)]=\\
C(m_D)+\int^r\dd r'\pdv{V(r',0)}{r'}e^{-m_Dr'} (m_Dr'+1),\label{ReV}
\end{multline}
\begin{multline}
\Im[V(r,m_D)]=\\
-\int^r\dd r'\pdv{V(r',0)}{r'}\frac{\sqrt{\pi }m_DT}{2} r'^2\\\times G^{2,1}_{1,3}\left(\left.\begin{array}{c}\frac{1}{2}\\\frac{1}{2},\frac{1}{2},-1\end{array}\right|\frac{1}{4}m_D^2r'^2\right).\label{ImV}
\end{multline}

The imaginary part may demand a regularization in the infrared, also required in the original work~\cite{Lafferty:2019jpr}. We will model it in a simpler way assuming that its dependence on $r$ is mild as discussed previously in this Section and, therefore, we take it as a constant.

\subsection{Specifics for the $X(3872)$}\label{sec1-3}
In this section we introduce the vacuum potential for the $X(3872)$ and how we apply the approach explained in the previous Subsection to calculate the in-medium potential.

\begin{figure}[!h]
\begin{center}
  \includegraphics[width=1\linewidth]{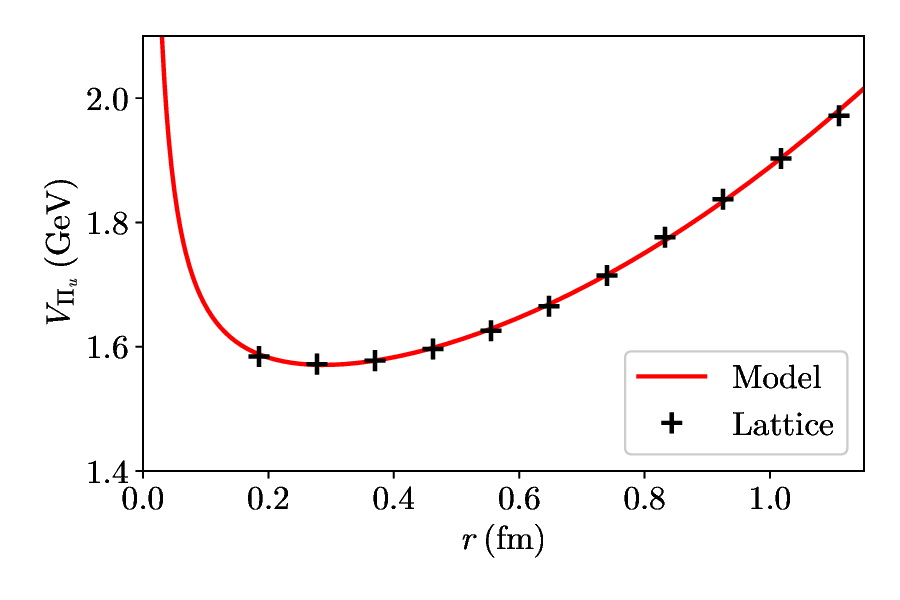}
\caption{The hybrid vacuum potential given in~\eqref{V0} with $A_{-1}=18.904\ \mathrm{MeV\cdot fm}$, $A_0=1471.953\ \mathrm{MeV}$ and $A_2=398.666\ \mathrm{MeV\cdot fm^{-2}}$, values extracted from \cite{Capitani:2018rox}.}\label{fig:V0}
\end{center}
\end{figure}




\subsubsection{The vacuum potential}
Recent lattice QCD computations of static potentials for hybrid mesons have shown that one state ($\Pi_u^-$) with quantum numbers $J^{PC}=1^{++}$ and a mass of roughly $4\ \mathrm{GeV}$ ($m_{\Pi_u^-}=4184\ \mathrm{MeV}$) arises in the spectrum~\cite{Capitani:2018rox}. We will take this state as the closest one to the $X(3872)$. Although the $X(3872)$ is often proposed as a tetraquark candidate, we consider that the potential between the two heavy valence quarks of the tetraquark resembles that of the hybrid since in both cases the valence quarks are influenced by light particles with color charge.


Physically, what we are assuming is that an ensemble of gluons behaves similarly to a light quark-antiquark pair with the same quantum numbers since in both cases it is possible to obtain octet and singlet representations.
With all the above in mind, we consider the vacuum potential to be the one of the hybrid $\Pi_u^-$
in the {\it singlet channel approximation}~\cite{Capitani:2018rox}
with the functional form given in~\eqref{V0} and shown in Fig.~\ref{fig:V0}, whose parameters
 for the $X(3872)$ will be discussed below.
 This is an educated guess for a model of the vacuum potential that establishes a simple starting point to obtain the in-medium one.


\subsubsection{The real part of the in-medium potential}
The real part of the in-medium potential can be computed using~\eqref{V0},~\eqref{ReV} and the prescription to determine $C(m_D)$ explained previously, to read
    \begin{multline}
    \Re[V(r,m_D)]=A_{-1}\left(m_D+\frac{e^{-m_Dr}}{r}\right)+A_0\\+A_2\left[\frac{6}{m_D^2}(1-e^{-m_Dr})-\left(2r^2+\frac{6r}{m_D}\right)e^{-m_Dr}\right].\label{ReVX}
\end{multline}

\begin{figure}[!h]
\begin{center}
  \includegraphics[width=1\linewidth]{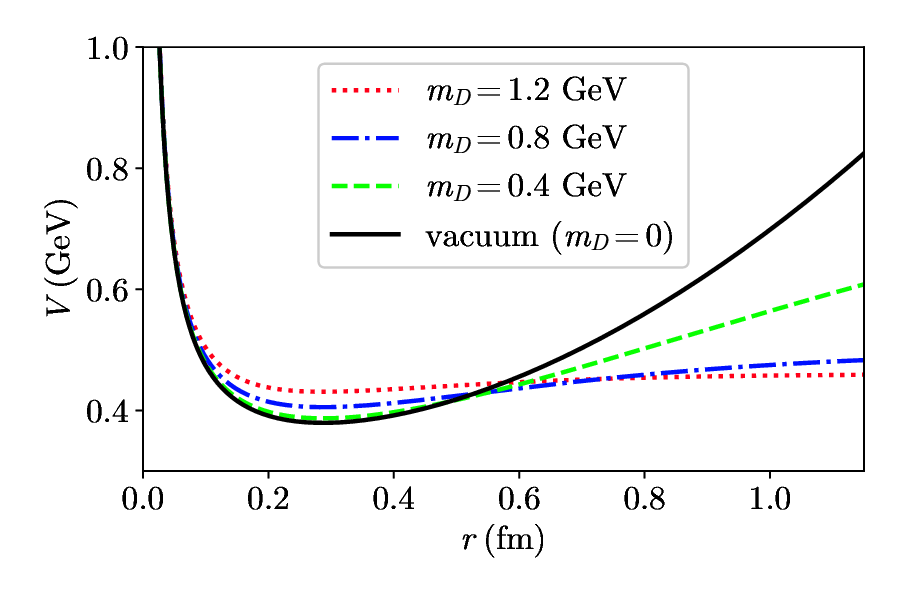}
\caption{The real part of the in-medium potential as given by~\eqref{ReVX} with $A_{-1}=18.904\ \mathrm{MeV\cdot fm}$, $A_0=280.724\ \mathrm{MeV}$ and $A_2=398.666\ \mathrm{MeV\cdot fm^{-2}}$.}\label{fig:Vm}
\end{center}
\end{figure}

Concerning the parameters in the vacuum potential for the $X(3872)$, we keep the ones for the Coulomb and quadratic terms ($A_{-1}$ and $A_2$) as given by the lattice computation for the hybrid, while we change the offset $A_0$ to reproduce the $X(3872)$ mass for the ground state through the Schr\"odinger equation. To do so, we define the binding energy such that, at $T=0$, $E=m_{X(3872)}-2m_c$ with $m_c=1.628\ \mathrm{GeV}$ (the value used in~\cite{Capitani:2018rox}).

We show the real part of the potential in Fig.~\ref{fig:Vm}, where one can see that it saturates at every $T>0$ to give
\begin{equation}
\lim_{r\to\infty}\Re[V(r,m_D)]=A_1m_D+A_0+\frac{6A_2}{m_D^2}\ .\label{lim}
\end{equation}
\subsubsection{The imaginary part}

For the imaginary part of the in-medium potential, we assume that it is a constant, as discussed previously. Here we provide the arguments that lead us to extract its thermal dependence.

The first argument is based on dimensional analysis. Given that the $T=0$ potential depends on three parameters ($A_{-1}$, $A_0$ and $A_2$), we expect that the decay width is the sum of two pieces, one proportional to $A_{-1}$ and another proportional to $A_2$, since $A_0$ is just an energy shift. Then, by dimensional analysis, we anticipate that the decay width goes like
\begin{equation}
    \Gamma=\alpha A_{-1}T+\beta A_2\frac{T}{m_D^3}\ , \label{gamma}
\end{equation}
where $\alpha$ and $\beta$ are dimensionless constants. The specific dependence with $T$ and $m_D$ is a consequence of the fact that the decay is dominated by relatively low energy modes (below the scale $T$). From the diagrammatic point of view, a decay width always implies a cut. In thermal field theory, cuts are related with the symmetric propagator which is proportional to $T$ for bosonic energy modes with energy much smaller than $T$. In fact, this structure can be seen already in~\eqref{eq:permitivity}. Therefore, the decay width must be proportional to $T$ and the remaining temperature dependence can only enter indirectly through $m_D$ and its power is determined by the dimensions of the constants\footnote{$A_{-1}$ is dimensionless and $A_2$ has dimensions of (energy)$^3$.}. 

A complementary argument is based on the application of the same approach that we used for the real part of the potential. This model must be adapted in two ways in order to apply it to the imaginary part of the potential. First, the formalism was originally formulated to treat the singlet case. However, we are now interested in a heavy quark-antiquark pair in the color octet configuration. We use the one-gluon exchange approximation to obtain how the color factors are modified when going from a singlet to an octet. In the singlet case, the decay width behaves as
\begin{equation}
    \Gamma_\mathrm{s}(r,m_D)=C_F(W(0,m_D)-W(r,m_D))\ ,
\end{equation}
being $W(r,m_D)$ the gluon propagator in coordinate space. In the octet case we have that
\begin{equation}
    \Gamma_\mathrm{o}(r,m_D)=C_FW(0,m_D)+\frac{1}{2N_c}W(r,m_D)\ .
\end{equation}
Since $W(r,m_D)$ vanishes as $r\to\infty$, at leading order, in the large $N_c$ limit, we get that
\begin{equation}
    \Gamma_\mathrm{o}(r,m_D)=\lim_{r\to\infty}\Gamma_\mathrm{s}(r,m_D)\ ,\label{decayO}
\end{equation}
where $\Gamma_\mathrm{s}(r,m_D)=\Im[V(r,m_D)]$ with the imaginary part of the potential as given in~\eqref{ImV}.

Now, by using~\eqref{ImV} and~\eqref{V0} we arrive at
\begin{multline}
    \Im[V(r,m_D)]=\\\frac{\sqrt{\pi } T}{m_D}\frac{A_{-1}}{r} G_{1,3}^{2,1}\left(\left.\begin{array}{c}
 \frac{3}{2} \\
 \frac{3}{2},\frac{3}{2},0 \\
\end{array}
\right|\frac{1}{4}m_D^2r^2\right)\\+4\frac{\sqrt{\pi } T}{m_D^3}A_2 G_{1,3}^{2,1}\left(\left.\begin{array}{c}
 \frac{3}{2} \\
 \frac{5}{2},\frac{5}{2},0 \\
\end{array}
\right|\frac{1}{4}m_D^2r^2\right)\ .
\end{multline}
Note that if we apply~\eqref{decayO} to the above equation we obtain a finite contribution from the first term of the right hand side while we have to deal with a divergence on the second term. The contribution from the first term gives $A_{-1}T$ which fixes $\alpha=1$ in~\eqref{gamma}. The limit of the second term, $\Im[V_2(r,m_D)]$, has to be regularized as $r\to\infty$. In order to do so we follow the approach in~\cite{Lafferty:2019jpr}.

\begin{figure}[!h]
\begin{center}
  \includegraphics[width=1\linewidth]{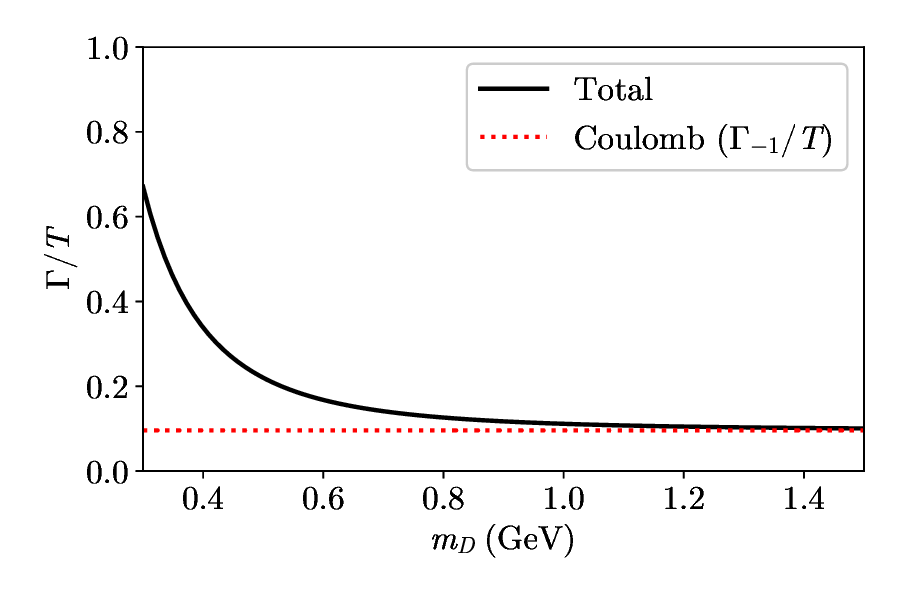}
\caption{The decay width scaled by the temperature vs. the Debye mass.}\label{fig:GT}
\end{center}
\end{figure}

From~\eqref{eq:permitivity} we can obtain
\begin{equation}
    \Im\varepsilon^{-1}(r,m_D)=-\frac{T}{2\pi}\int_0^\infty\dd p\frac{p^2m_D^2\sin{pr}}{r(p^2+m_D^2)^2}\ ,
\end{equation}
so
\begin{multline}
\Im[V_2(r,m_D)]=\\
-4A_2Tm_D^2\int_0^\infty\dd p\frac{(3-p^2r^2)\sin{pr}-3pr\cos{pr}}{p^2(p^2+m_D^2)^2}
\end{multline}
which can be written, through the change of variables $u=p/m_D$ and $x=m_Dr$, as
\begin{multline}
\Im[V_2(r,m_D)]=\\-\frac{4A_2T}{m_D^3}\int_0^\infty\dd u\frac{(3-u^2x^2)\sin{ux}-3ux\cos{ux}}{u^2(u^2+1)^2}\ .
\end{multline}
The integral above is not defined in the limit $x\to\infty$. To regularize it we introduce the function
\begin{multline}
    \chi_n(x,\Delta)=\\4\int_0^\infty\dd u\frac{(3-u^2x^2)\sin{ux}-3ux\cos{ux}}{u^{2-n}(u^2+\Delta^2)^{n/2}(u^2+1)^2}\ ,\label{chin}
\end{multline}
that for $x\to\infty$ reads 
\begin{equation}
    \chi_n(x,\Delta)=\frac{6}{\Delta^n}\int_0^\infty\dd u\frac{\sin{ux}}{u^{2-n}}
\end{equation}
which diverges if $n\leq0$ and cancels if $n\geq2$ but is finite for $n=1$, so 
\begin{equation}
    \lim_{x\to\infty}\chi_n(x,\Delta)=\left\{\begin{array}{ll}
\infty,&n\leq0\\
6\pi/\Delta,&n=1\\
0,&n\geq2
\end{array}\right.\ .
\end{equation}
The only non trivial choice is $n=1$, so the limit is
\begin{equation}
    \lim_{r\to\infty}\Im[V_2(r,m_D)]=\frac{6\pi A_2T}{m_D^3\Delta}
\end{equation}
and, therefore,
\begin{multline}
    \Gamma_\mathrm{o}(r,m_D)=A_{-1}T+\frac{A_2T}{m_D^3}\frac{6\pi}{\Delta}=\\\Gamma_{-1}(m_D)+\Gamma_2(m_D)\ .\label{gammaT}
\end{multline}
We can relate $\beta$ to $\Delta$ by comparing with~\eqref{gamma}, obtaining $\beta=6\pi/\Delta$. We fix the value of $\Delta$ in an analogous way as in~\cite{Lafferty:2019jpr} to be $\Delta=6\pi$. In any case, it is clear that, for large temperatures, the Coulomb contribution, $\Gamma_{-1}(m_D)$, dominates over the quadratic contribution, $\Gamma_2(m_D)$, in the decay width as shown in Fig.~\ref{fig:GT}.

\section{Results}\label{sec2}

\subsection{Dissociation temperature}
\begin{figure}[!h]
\begin{center}
  \includegraphics[width=1\linewidth]{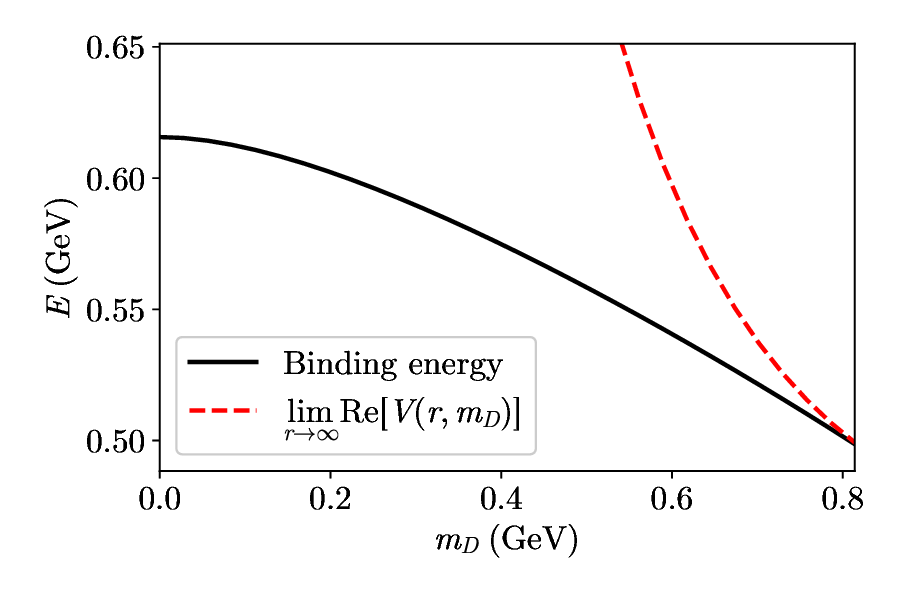}
\caption{Binding energy and the large $r$ real potential as given by~\eqref{lim} vs. Debye mass.}\label{fig:EligLIM}
\end{center}
\end{figure}

\begin{figure}[!h]
\begin{center}
  \includegraphics[width=1\linewidth]{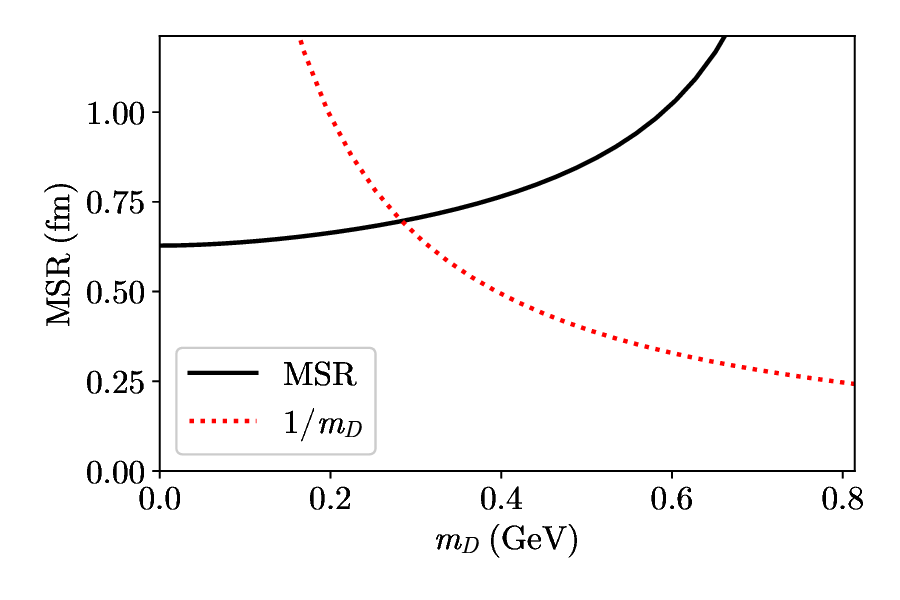}
\caption{Mean square radius, $\sqrt{\langle r^2\rangle}$, and the screening length, $1/m_D$, vs. Debye mass.}\label{fig:MSR}
\end{center}
\end{figure}

Let us now discuss the dissociation temperature, defined as the temperature above which the Schr\"odinger equation no longer admits bound states. Since the imaginary part of the potential is a constant, it does not affect the solution. In Fig.~\ref{fig:EligLIM} the behavior of the binding energy compared to~\eqref{lim} versus the Debye mass is shown. The binding energy has been computed numerically using the Numerov method and the previously mentioned value of the heavy quark mass, $m_c=1.628\ \textrm{GeV}$. We observe that there are no bound states solutions with the potential we are using for Debye masses above $m_D\sim 700\ \mathrm{MeV}$. This value agrees approximately with the one at which the $\lim_{r\to\infty}\Re[V(r,m_D)]$ equals the binding energy.
As seen in Fig.~\ref{fig:MSR}, this temperature coincides with a sharp increase of the mean square radius (MSR), and that this radius has a very mild increase with temperature up to the point in which it is comparable with the screening length (the inverse of the Debye mass).

In summary, we obtain a dissociation temperature $T\sim 350\,\mathrm{MeV}$. To get this approximate value we apply the relation between $T$ and $m_D$ given in~\cite{Lafferty:2019jpr} where an HTL interpolation formula is described.

\subsection{The survival probability}

The survival probability is exponentially related to the decay rate by
\begin{equation}
    S(t)=\exp[-\int_{t_0}^{t}\dd\tau\Gamma(T(\tau),\tau)]\ ,
\end{equation}
where the dependence of the temperature with the time is explicit. Let us note that since in our case the decay width does not depend on $r$, the previous formula is exact.

\begin{figure}[!h]
\begin{center}
  \includegraphics[width=1\linewidth]{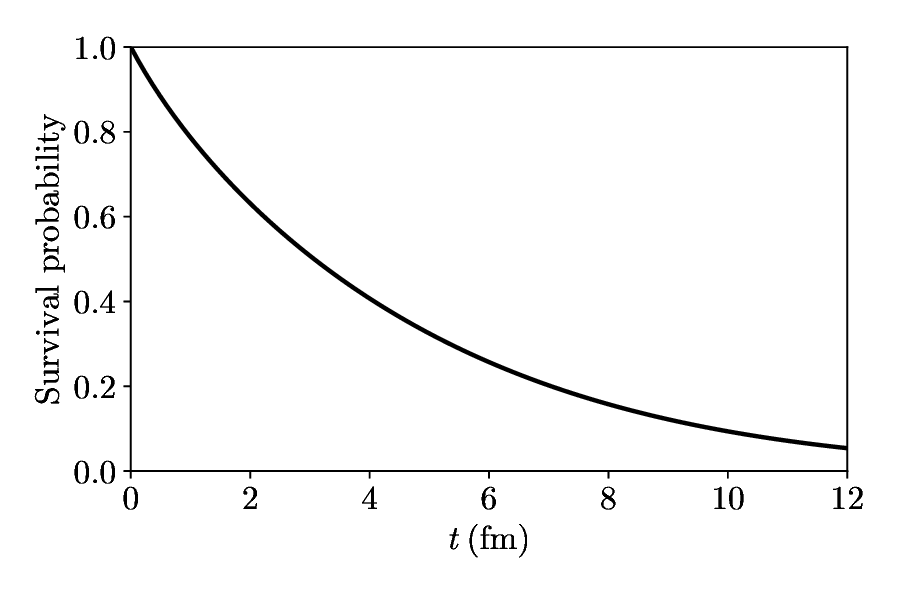}
\caption{Survival probability vs. time.}\label{fig:S}
\end{center}
\end{figure}

As a first approximation, we assume Bjorken expansion~\cite{Bjorken:1982qr}. This implies that $tT^3$ behaves as a constant. Using~\eqref{gammaT} and assuming a linear relation between $m_D$ and $T$ (a result from perturbation theory, modulo small logarithmic corrections) for $\Gamma(T)$, we can separate 
$S(t)=S_{-1}(t)S_2(t)$ where
\begin{eqnarray}
    S_{-1}(t)=&\exp[-\int_{t_0}^{t}\dd\tau\Gamma_{-1}(T(\tau),\tau)]\ ,\\
    S_2(t)=&\exp[-\int_{t_0}^{t}\dd\tau\Gamma_2(T(\tau),\tau)]\ .
\end{eqnarray}
Since $tT^3$ is a constant, so is $t^{-n/3}\Gamma_n$, leading to
\begin{eqnarray}
    S_{-1}(t)=&\exp{-\frac{3t_0\Gamma_{-1}(T_0)}{2}\left[\left(\frac{t}{t_0}\right)^{2/3}-1\right]}\ ,\\
    S_2(t)=&\exp{-\frac{3t_0\Gamma_2(T_0)}{5}\left[\left(\frac{t}{t_0}\right)^{5/3}-1\right]}\ .
\end{eqnarray}
Our result for the total survival probability $S(t)$ is shown in Fig.~\ref{fig:S} where we have used $t_0=0.6\ \mathrm{fm}$ and $T_0=500\ \mathrm{MeV}$ ($m_D(T_0)\simeq1.15\ \mathrm{GeV}$ \cite{Lafferty:2019jpr}), which correspond to commonly used values at LHC energies.

\begin{figure}[!h]
\begin{center}
  \includegraphics[width=1\linewidth]{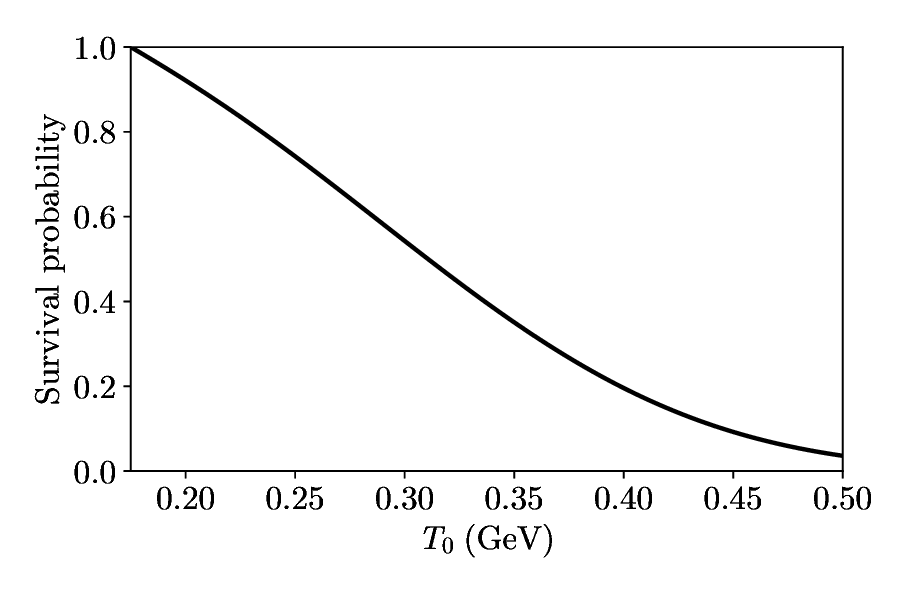}
\caption{Survival probability vs. initial temperature.}\label{fig:ST}
\end{center}
\end{figure}

In Fig.~\ref{fig:ST} we show the dependence of the survival probability computed at the time when the temperature has undergone a Bjorken expansion down to the phase transition value $T_c=175\ \mathrm{MeV}$, versus the initial temperature $T_0$.

Our combined results for the dissociation temperature and the survival probability provide the following qualitative picture:
\begin{itemize}
    \item If the initial temperature is larger than $350\ \mathrm{MeV}$, the $X(3872)$ does not form in the medium. It may reappear later when the system cools down as a consequence of uncorrelated recombination, an effect that we have not considered.
    \item If the initial temperature is smaller than $350\ \mathrm{MeV}$, the system does form but has a finite probability to decay. As seen in Fig.~\ref{fig:ST}, this decay probability is small for initial temperatures below $T\sim 250\ \mathrm{MeV}$, but becomes sizable for higher temperatures.
\end{itemize}
    Therefore, assuming a tetraquark nature for the $X(3872)$, we conclude that  the dominant dissociation mechanism is Debye screening for central heavy-ion collisions, while for peripheral ones the finite width plays a significant role.

\section{Conclusions}\label{sec3}

In this manuscript we address the problem of the behavior of exotic states in heavy-ion collisions. We focus on the $X(3872)$ that we consider to be a tetraquark.
We propose a lattice-inspired potential for the $X(3872)$, which we have extended for finite temperatures. Note that this approach could be applied to other exotic states. It could also be applied to other internal structures like a molecule, provided that the Born-Oppenheimer approximation is applicable, e.g., the exotic state taken as a bound state of two light-heavy singlets, or to any potential  given as a function of a single distance.

Our derivation of the real part of the potential relies on the approach in~\cite{Lafferty:2019jpr}, using as input the potential for hybrids obtained on the lattice~\cite{Capitani:2018rox}.
Using this real part of the potential in the Schr\"odinger equation, we have estimated the dissociation temperature to be around $350\ \mathrm{MeV}$.

We argue, based on large $N_c$ considerations, that the imaginary part of the potential is a mild function of $r$, and we take it to be a constant. Using this imaginary part, we have computed the survival probability assuming that the system undergoes a Bjorken expansion.

The results obtained here suggest that for central heavy-ion collisions the dominant mechanism for the in-medium evolution of the $X(3872)$ is Debye screening, while for peripheral collisions  the finite width plays a sizable role.

For the future, we leave the inclusion of this potential in a Lindblad equation, which would include the effect of correlated recombination, although this is expected to be a small effect as it is seen for bottomonium states~\cite{Brambilla:2020qwo}.

We have not considered uncorrelated recombination, i.e., that coming from different heavy quark pairs, either. We note that our approach could easily accommodate it, as our imaginary part is a constant. Finally, phenomenological applications are also left for future work.

\bmhead{Acknowledgments}

We thank Alexander Rothkopf and Joan Soto for useful discussions. We have received financial support from Xunta de Galicia (Centro singular de investigación de Galicia accreditation 2019-2022, ref. ED421G-2019/05), by European Union ERDF, by the ``María de Maeztu'' Units of Excellence program MDM2016-0692, and by the Spanish Research State Agency under project PID2020-119632GBI00. This work has received funding from the European Research Council project ERC-2018-ADG-835105 YoctoLHC and from the European Union’s Horizon 2020 research and innovation programme under grant agreement No. 824093. The work of MAE has been supported by the Maria de Maetzu excellence program under project CEX2019-000918-M, by projects PID2019-105614GB-C21 and PID2022-136224NB-C21 funded by MCIN/AEI/10.13039/501100011033, and by grant 2021-SGR-249 of Generalitat de Catalunya. VLP has been supported by Xunta de Galicia under project ED481A 2022/286.

\bibliographystyle{bst/sn-mathphys}
\bibliography{sn-bibliography.bib}

\end{document}